\documentclass[sigconf,screen]{acmart}
\renewcommand\footnotetextcopyrightpermission[1]{} % removes footnote with conference information in first column
\usepackage{booktabs}
\usepackage{multirow}
\usepackage{makecell}
\usepackage{subfigure}
\usepackage[misc]{ifsym} 
\usepackage{balance}
\usepackage{listings}
\usepackage{color, xcolor}
\usepackage{algorithm2e}
\usepackage{diagbox}
\usepackage{tcolorbox}
\usepackage{stfloats}
\usepackage{enumitem}
\usepackage{ulem}
\AtBeginDocument{%
  \providecommand\BibTeX{{%
    \normalfont B\kern-0.5em{\scshape i\kern-0.25em b}\kern-0.8em\TeX}}}

\setcopyright{acmcopyright}
\acmYear{2026}
\copyrightyear{2025}
\setcopyright{acmlicensed}

\begin{document}
\begin{sloppypar}
	
	\author{Lingzhe Zhang}
	\affiliation{%
		\institution{Peking University}
		\city{Beijing}
		\country{China}}
	\email{zhang.lingzhe@stu.pku.edu.cn}
	
	\author{Tong Jia$^{\ast}$}
	\affiliation{%
		\institution{Peking University}
		\city{Beijing}
		\country{China}}
	\email{jia.tong@pku.edu.cn}
	
	\author{Weijie Hong}
	\affiliation{%
		\institution{Peking University}
		\city{Beijing}
		\country{China}}
	\email{hongwj@stu.pku.edu.cn}
	
	\author{Mingyu Wang}
	\affiliation{%
		\institution{Peking University}
		\city{Beijing}
		\country{China}}
	\email{mingyuwang25@stu.pku.edu.cn}
	
	\author{Chiming Duan}
	\affiliation{%
		\institution{Peking University}
		\city{Beijing}
		\country{China}}
	\email{duanchiming@stu.pku.edu.cn}
	
	\author{Minghua He}
	\affiliation{%
		\institution{Peking University}
		\city{Beijing}
		\country{China}}
	\email{hemh2120@stu.pku.edu.cn}
	
	\author{Rongqian Wang}
	\affiliation{%
		\institution{Huawei Technologies Co., Ltd.}
		\city{Beijing}
		\country{China}}
	\email{wangrongqian2@huawei.com}
	
	\author{Xi Peng}
	\affiliation{%
		\institution{Huawei Technologies Co., Ltd.}
		\city{Hong Kong SAR}
		\country{China}}
	\email{pancy.pengxi@huawei.com}
	
	\author{Meiling Wang}
	\affiliation{%
		\institution{Huawei Technologies Co., Ltd.}
		\city{Shenzhen}
		\country{China}}
	\email{wangmeiling17@huawei.com}
	
	\author{Gong Zhang}
	\affiliation{%
		\institution{Huawei Technologies Co., Ltd.}
		\city{Shenzhen}
		\country{China}}
	\email{nicholas.zhang@huawei.com}
	
	\author{Renhai Chen}
	\affiliation{%
		\institution{Huawei Technologies Co., Ltd.}
		\city{Beijing}
		\country{China}}
	\email{chenrenhai@huawei.com}
	
	\author{Ying Li$^{\ast}$}
	\affiliation{%
		\institution{Peking University}
		\city{Beijing}
		\country{China}}
	\email{li.ying@pku.edu.cn}
	
	\renewcommand{\shortauthors}{Lingzhe Zhang et al.}

\title[RuntimeSlicer: Towards Generalizable Unified Runtime State Representation for Failure Management]{RuntimeSlicer: Towards Generalizable Unified Runtime State Representation for Failure Management}

\begin{abstract}
	Modern software systems operate at unprecedented scale and complexity, where effective failure management is critical yet increasingly challenging. Metrics, traces, and logs provide complementary views of system runtime behavior, but existing failure management approaches typically rely on task-oriented pipelines that tightly couple modality-specific preprocessing, representation learning, and downstream models, resulting in limited generalization across tasks and systems. To fill this gap, we propose RuntimeSlicer, a unified runtime state representation model towards generalizable failure management. RuntimeSlicer pre-trains a task-agnostic representation model that directly encodes metrics, traces, and logs into a single, aligned system-state embedding capturing the holistic runtime condition of the system. To train RuntimeSlicer, we introduce Unified Runtime Contrastive Learning, which integrates heterogeneous training data sources and optimizes complementary objectives for cross-modality alignment and temporal consistency. Building upon the learned system-state embeddings, we further propose State-Aware Task-Oriented Tuning, which performs unsupervised partitioning of runtime states and enables state-conditioned adaptation for downstream tasks. This design allows lightweight task-oriented models to be trained on top of the unified embedding without redesigning modality-specific encoders or preprocessing pipelines. Preliminary experiments on the AIOps 2022 dataset demonstrate the feasibility and effectiveness of RuntimeSlicer for system state modeling and failure management tasks.
\end{abstract}




\maketitle

\section{Introduction}

Modern software systems continue to evolve at an unprecedented scale and level of complexity. From e-commerce platforms and social media services to financial trading systems and cloud-native infrastructures, today’s software systems routinely support billions of users and execute business-critical workloads under highly dynamic conditions.

However, the widespread adoption of microservice architectures, containerized deployments, and elastic scaling mechanisms has significantly increased the dynamism, heterogeneity, and unpredictability of runtime environments. As a result, large-scale software systems experience failures more frequently, leading to substantial operational disruptions and financial losses~\cite{elliot2014devops, zhang2025survey}. According to an ITIC report, unplanned IT downtime in production environments costs large enterprises over one million U.S. dollars per hour on average~\cite{ITICCostofDowntime2024}. These challenges highlight the critical need for timely failure management, including anomaly detection, failure localization and failure classification.

System traces, metrics, and logs capture rich information about the runtime behavior of software systems, recording both the internal states of executing components and the critical events occurring during system operation. As such, they constitute data sources for failure management, providing insights into normal system behavior as well as deviations that may indicate emerging failures. 

\vspace{-0.8em}
\begin{figure}[htbp]
	\centering
	\includegraphics[width=1\linewidth]{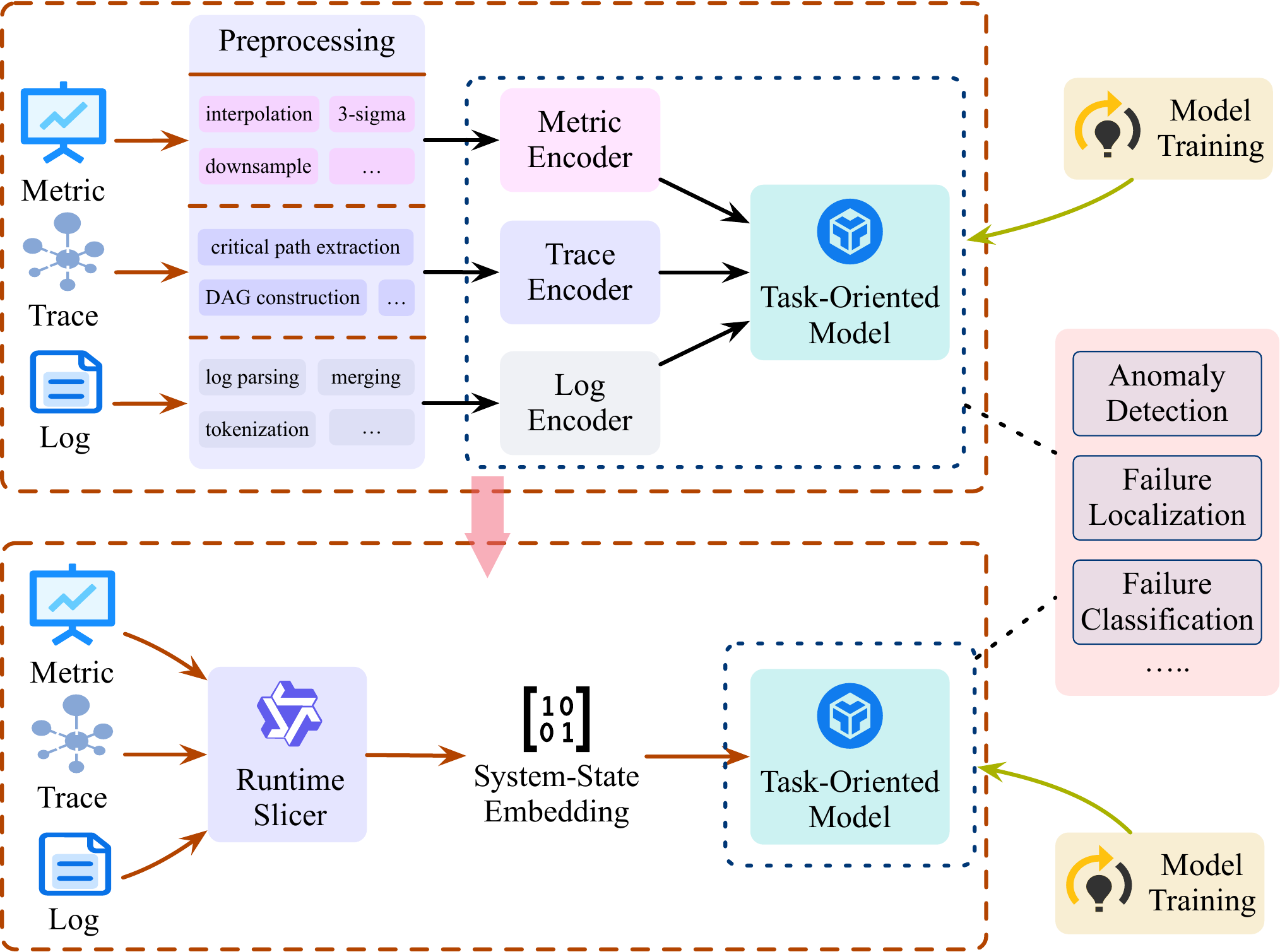}
	\vspace{-1.2em}
	\caption{Comparison between conventional failure management training pipelines and the RuntimeSlicer-enabled unified training workflow.}
	\label{fig: intro-example}
	\vspace{-0.8em}
\end{figure}

A large body of prior work has explored failure management techniques based on these runtime data modalities. Metric-based approaches analyze key indicators of system performance and resource utilization—such as latency, throughput, and CPU or memory usage—to detect anomalies or infer potential failure causes~\cite{liu2025ora, yu2021microrank, lin2024root, rasul2023lag, dong2024simmtm, liao2025timegpt, zhang2025microremed, liu2025aaad}.  Log-based methods leverage system logs as sequences of structured or semi-structured events, combining temporal patterns and semantic information to uncover abnormal behaviors and failure signatures~\cite{sui2023logkg, li2022swisslog, zhang2025scalalog, zhang2025xraglog, zhang2024multivariate, zhang2024reducing, zhang2025log, zhang2025adaptive, duan2025logaction, hong2025cslparser, he2025walk, he2025united}. Trace-based approaches focus on service invocation relationships and request propagation paths across distributed components, and are commonly applied to failure localization and failure classification tasks in microservice systems~\cite{li2022microsketch, yu2023tracerank, gan2021sage, cai2021tracemodel, zhang2022crisp, zhang2024trace}. More recently, multimodal approaches attempt to jointly leverage logs, metrics, and traces to improve failure detection and diagnosis by fusing complementary runtime signals from different data sources~\cite{liu2024uac, zhang2025thinkfl, zhang2025agentfm, lee2023eadro, zhang2023robust, yu2023nezha, sun2025interpretable, sun2024art, zhang2026agentic, zhang2026hypothesize, huang2025uda}. 

The effectiveness of existing failure management methods has been demonstrated. However, as illustrated in Figure~\ref{fig: intro-example}, current approaches still suffer from fundamental limitations. Specifically, most existing methods adopt a task-oriented pipeline: metrics, traces, and logs are first processed through modality-specific preprocessing procedures, followed by separate encoders tailored to each data modality. The resulting representations are then fused and fed into task-specific models, such as anomaly detection, failure localization, or failure classification models. During training, the modality encoders and task-oriented models are optimized jointly in an end-to-end manner. This tightly coupled design causes failure-related patterns to be implicitly entangled with task objectives and encoder architectures, making the learned representations highly dependent on specific tasks, datasets, and system configurations. As a result, the extracted features are difficult to reuse across tasks or environments, leading to limited generalization when system behaviors, workloads, or failure patterns change.

\begin{figure*}[htbp]
	\centering
	\includegraphics[width=1\linewidth]{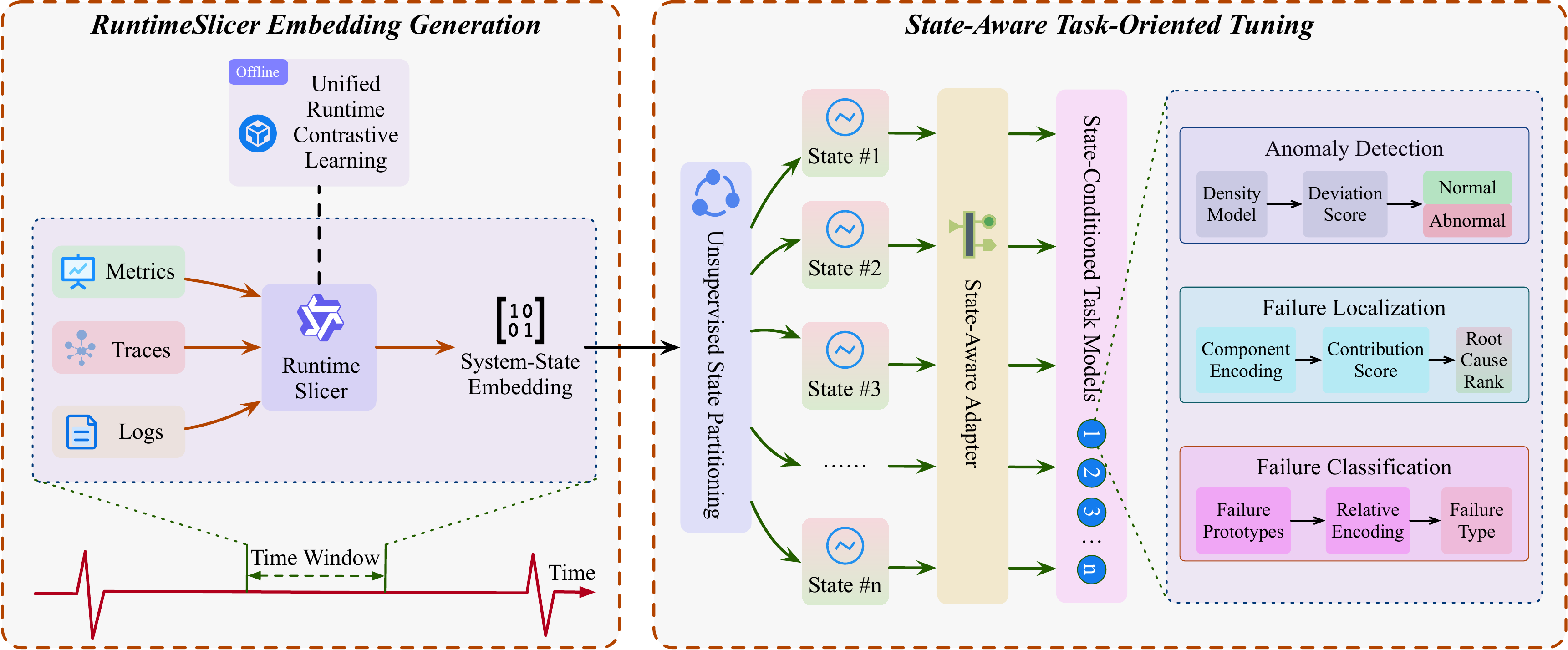}
	\vspace{-1.2em}
	\caption{Pipeline of RuntimeSlicer for failure management. The left side illustrates how RuntimeSlicer, trained via Unified Runtime Contrastive Learning, ingests metrics, traces, and logs to produce a system-state embedding. The right side shows how this embedding is leveraged for State-Aware Task-Oriented Tuning of downstream failure management models.}
	\label{fig: workflow}
\end{figure*}

To fill this gap, we propose \textbf{RuntimeSlicer}, a \uline{unified runtime state representation model towards generalizable failure management}. RuntimeSlicer pre-trains a \uline{task-agnostic} representation model that directly ingests metrics, traces, and logs, and encodes them into a single, aligned system-state embedding capturing the holistic runtime condition of the system.

Unlike conventional task-oriented pipelines that jointly train modality-specific encoders and downstream models, RuntimeSlicer decouples representation learning from failure management tasks. Once trained, RuntimeSlicer can be applied to different systems to produce system-state embeddings, on top of which lightweight task-oriented models—such as anomaly detection, failure localization, and failure classification—can be \uline{trained without redesigning preprocessing pipelines or re-training modality encoders}.

To train RuntimeSlicer, we introduce \textbf{Unified Runtime Contrastive Learning}, a representation learning framework that integrates diverse training data sources, including labeled datasets, runtime collection from live systems, and controlled failure injection. The training objective combines multiple complementary loss functions, including modal consistency loss for cross-modality alignment, temporal consistency loss for preserving runtime continuity, and an optional weak anomaly loss to incorporate coarse-grained failure signals.

Building upon the learned system-state embeddings, we further observe that system states are naturally structured and recurrent, corresponding to distinct runtime conditions such as workload levels, traffic patterns. Motivated by this observation, we propose \textbf{State-Aware Task-Oriented Tuning}, which first performs unsupervised state partitioning over system-state embeddings to identify latent runtime states. For each identified state, RuntimeSlicer employs a state-aware adaptation mechanism to train state-conditioned task models, enabling downstream failure management tasks to adapt to heterogeneous runtime conditions.

We conduct preliminary experiments on the AIOps 2022 dataset~\cite{aiops2022championship}, which is collected from a mature microservices-based e-commerce system, to evaluate the effectiveness of RuntimeSlicer in distinguishing system runtime states and supporting failure management tasks.

\section{Methodology}

RuntimeSlicer-enabled failure management is organized into two main stages: system-state embedding generation and state-aware task-oriented tuning. As shown in Figure~\ref{fig: workflow}, the left part of the pipeline illustrates the embedding generation process, where metrics, traces, and logs collected within the same time window are jointly fed into a pre-trained RuntimeSlicer model. RuntimeSlicer, trained via Unified Runtime Contrastive Learning, encodes these runtime signals into a compact system-state embedding.

Once the system-state embedding is obtained, it can be used as input to the State-Aware Task-Oriented Tuning stage, which supports downstream failure management models. It is worth noting that the primary goal of this paper is to pre-train a generalizable runtime state representation through RuntimeSlicer. The proposed state-aware task-oriented tuning serves as one possible instantiation demonstrating how the learned embedding can be exploited, while the representation itself is compatible with a wide range of downstream tasks and model designs.

\subsection{Unified Runtime Contrastive Learning}

\begin{figure*}[htbp]
	\centering
	\includegraphics[width=1\linewidth]{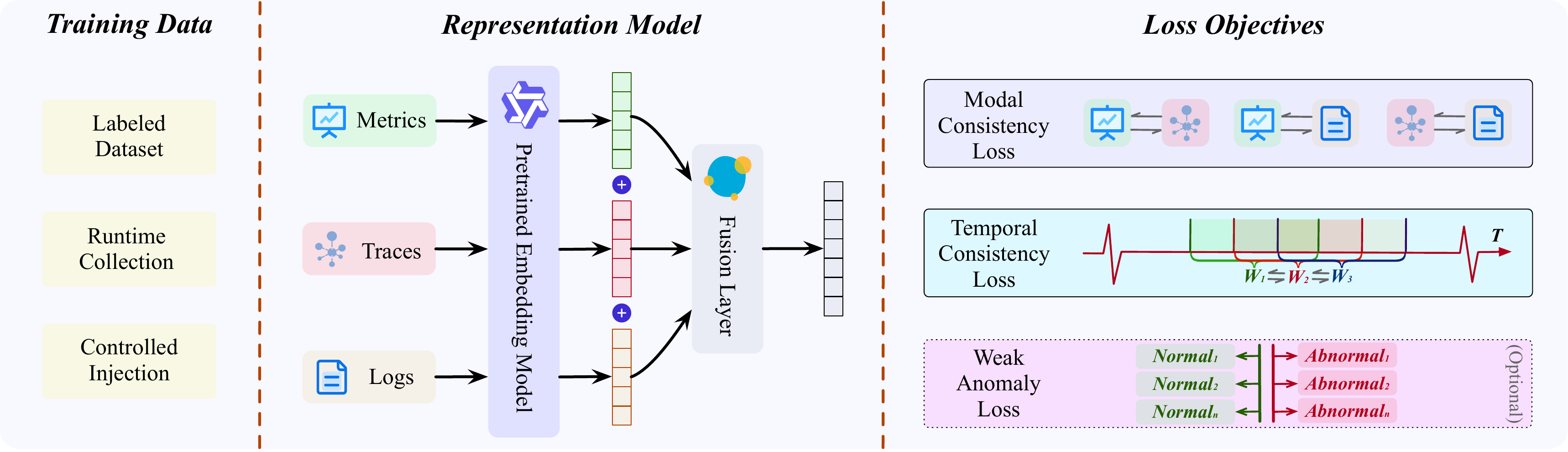}
	\vspace{-1.2em}
	\caption{Training Pipeline of Unified Runtime Contrastive Learning}
	\label{fig: training}
\end{figure*}

Unified Runtime Contrastive Learning is designed to learn a unified and task-agnostic representation of system runtime states from heterogeneous observability data. Unlike task-oriented training paradigms that entangle representation learning with specific failure management objectives, our goal is to pre-train a representation model that captures the intrinsic structure of runtime behaviors and can be reused across different systems and downstream tasks. As shown in Figure~\ref{fig: training}, Unified Runtime Contrastive Learning consists of three key components: training data construction, representation model, and loss objectives.

\textbf{Training Data Construction.}
To support robust and scalable representation learning, we leverage three complementary sources of training data:

\uline{(1) Labeled Datasets.} We utilize publicly available datasets, such as AIOps 2022~\cite{aiops2022championship} and ART~\cite{sun2024art}, which provide synchronized metrics, traces, and logs along with failure annotations. These datasets offer explicit supervision and serve as a reference for learning failure-aware runtime representations.

\uline{(2) Runtime Collection.} We deploy a set of open-source distributed software systems, including Train-Ticket~\cite{zhou2018fault} and Online Boutique~\cite{google2025onlineboutique}, under diverse workload configurations. During execution, we continuously collect metrics, traces, and logs generated by the running systems without requiring manual labeling. This data source enables RuntimeSlicer to capture the intrinsic structure and natural variability of system runtime behaviors.

\uline{(3) Controlled Injection.} Building upon the runtime collection setup, we further introduce a limited amount of fault data through controlled failure injection using Chaos Mesh. This data source is intentionally kept small, as learning a unified runtime state representation does not fundamentally rely on explicit anomaly labels. Instead, injected failures are used to provide weak supervisory signals that enhance the model’s ability to distinguish coarse-grained abnormal and normal states.

\textbf{Representation Model.}
RuntimeSlicer is instantiated by integrating a shared pre-trained embedding backbone with minimal architectural adaptation for unified multimodal representation. Given a time window $t$, we denote a multimodal runtime observation as Equation~\ref{eq:input-data}.

\begin{equation}
	x_t = (x_t^{M}, x_t^{T}, x_t^{L})
	\label{eq:input-data}
\end{equation}

In the equation, $x_t^{M}$, $x_t^{T}$, and $x_t^{L}$ correspond to metrics, traces, and logs respectively. RuntimeSlicer adopts a unified pre-trained embedding backbone $f_{\theta}$ (Qwen3-Embedding-0.6B in our implementation), shared across modalities, to encode each input into a latent semantic embedding, as illustrated in Equation~\ref{eq:input-embedding}.

\begin{equation}
	e_t^{M} = f_{\theta}(x_t^{M}), \quad
	e_t^{T} = f_{\theta}(x_t^{T}), \quad
	e_t^{L} = f_{\theta}(x_t^{L}).
	\label{eq:input-embedding}
\end{equation}

Unlike conventional architectures that design modality-specific encoders independently, RuntimeSlicer leverages a shared linguistic–statistical embedding prior to ensure that heterogeneous signals are mapped into a unified embedding space, enabling natural comparability and modality-aligned reasoning.

The resulting modality embeddings are aggregated via a lightweight fusion layer $g_{\phi}$, implemented as a shallow multi-layer perceptron (MLP) with optional gating as Equation~\ref{eq:output-embedding}.

\begin{equation}
	z_t = g_{\phi}\left( \mathrm{Concat}(e_t^{M}, e_t^{T}, e_t^{L}) \right)
	\label{eq:output-embedding}
\end{equation}

The fusion layer finally producs a compact system-state embedding $z_t \in \mathbb{R}^{d}$ that summarizes the holistic runtime condition of the system, reflecting workload fluctuations, temporal dependencies, service interactions, and emergent failures.

\textbf{Loss Objectives.}
Unified Runtime Contrastive Learning jointly enforces three complementary constraints to shape the system-state embedding space.

\uline{(1) Modal Consistency Loss.}
To ensure that metrics, traces, and logs reflect a coherent system condition within the same time window, we align modality-specific embeddings using an InfoNCE-style contrastive term. For a batch of time windows $\{t_i\}_{i=1}^B$, let $z^{M}_i, z^{T}_i, z^{L}_i$ denote metric-, trace-, and log-level embeddings respectively. The loss aligns each modality pair by treating same-window representations as positives and others as negatives as Equation~\ref{eq:modal-loss}, where $\mathrm{sim}(\cdot)$ denotes cosine similarity and $\tau$ is a temperature parameter. Losses for $(T,L)$ and $(M,L)$ pairs are computed analogously.

\begin{equation}
	\mathcal{L}_{\text{modal}}
	=
	-\frac{1}{B}
	\sum_{i=1}^{B}
	\log
	\frac{
		\exp\big( \mathrm{sim}(z^{T}_i, z^{M}_i) / \tau \big)
	}{
		\sum_{j=1}^{B}
		\exp\big( \mathrm{sim}(z^{T}_i, z^{M}_j) / \tau \big)
	}
	\label{eq:modal-loss}
\end{equation}

\uline{(2) Temporal Consistency Loss.}
Runtime states exhibit smooth temporal evolution, where adjacent time windows sharing similar runtime spans are expected to remain semantically close in the embedding space. Let $s_i$ be the state embedding of window $t_i$, and $\omega_{ij}\in[0,1]$ denote their temporal-overlap ratio, which softly weights the expected similarity as Equation~\ref{eq:temp-loss}, where $\delta$ is a slack margin and $Z$ is a normalization constant. This formulation naturally induces smoothness for temporally coherent runtime periods while avoiding excessively forcing embeddings of unrelated windows to collapse, thereby preserving meaningful separation when temporal evidence suggests divergence.

\begin{equation}
	\mathcal{L}_{\text{temp}}
	=
	\frac{1}{Z}
	\sum_{i\neq j}
	\omega_{ij}\,
	\max\big( \delta - \mathrm{sim}(s_i, s_j), 0 \big)
	\label{eq:temp-loss}
\end{equation}

\uline{(3) Weak Anomaly Separation Loss (Optional).}
When coarse anomaly labels exist, we introduce a light constraint to prevent abnormal states from becoming overly similar to normal ones. Let $\mathcal{N}$ and $\mathcal{A}$ denote normal and abnormal sets of embeddings as Equation~\ref{eq:anom-loss}, where $\gamma$ controls the maximum permitted similarity. Notably, this term does not cluster anomalies, which only avoids accidental entanglement between normal and abnormal states.

\begin{equation}
	\mathcal{L}_{\text{anom}}
	=
	\mathbb{E}_{s_n \in \mathcal{N},\, s_a \in \mathcal{A}}
	\big[
	\max\big(\mathrm{sim}(s_n, s_a) - \gamma, 0\big)
	\big]
	\label{eq:anom-loss}
\end{equation}

\subsection{State-Aware Task-Oriented Tuning}

Since RuntimeSlicer fundamentally captures the underlying \emph{system state}, we further enable downstream failure management to become state-adaptive. Let $\mathcal{S}=\{s_1,\dots,s_N\}$ denote the learned system-state embeddings. We first perform unsupervised state partitioning as shown in Equation~\ref{eq:state-partitioning}, identifying latent runtime regimes such as workload tiers, traffic conditions, and resource-pressure states:

\begin{equation}
	\mathcal{C} = \{C_1,\dots,C_K\}, \quad 
	C_k = \{ s_i \in \mathcal{S} \mid \mathrm{assign}(s_i)=k \}.
	\label{eq:state-partitioning}
\end{equation}

To operationalize such structure, we introduce \uline{State-Aware Adapter}, a lightweight module that conditions downstream task models on cluster-specific contextual information. For a downstream failure-management task $\mathcal{T}$, we derive a set of \uline{State-Conditioned Task Models} $\{\theta_1,\dots,\theta_K\}$, each tuned on samples belonging to one state cluster as Equation~\ref{eq:state-model} which yields state-specialized models while maintaining a shared global backbone.

\begin{equation}
	\theta_k
	=
	\arg\min_{\theta}
	\;
	\mathbb{E}_{s_i \in C_k}
	\big[
	\mathcal{L}_{\mathcal{T}}(x_i; \theta, s_i)
	\big]
	\label{eq:state-model}
\end{equation}

For different downstream tasks, RuntimeSlicer instantiates the above mechanism as follows: \uline{Anomaly Detection}—state-conditioned density models estimate deviations from typical behavior to produce anomaly scores; \uline{Failure Localization}—state-aware component encodings yield contribution scores for root-cause ranking; and \uline{Failure Classification}—state-specific prototypes enable relative encoding–based classification within each state regime.

\section{Preliminary Evaluation}

To evaluate RuntimeSlicer, we conduct a preliminary study on the AIOps 2022 dataset to assess its feasibility.

\begin{figure}[htbp]
	\centering
	\subfigure[Qwen3-Embedding-0.6B]{
		\begin{minipage}{0.47\linewidth}
			\centering
			\includegraphics[width=\textwidth]{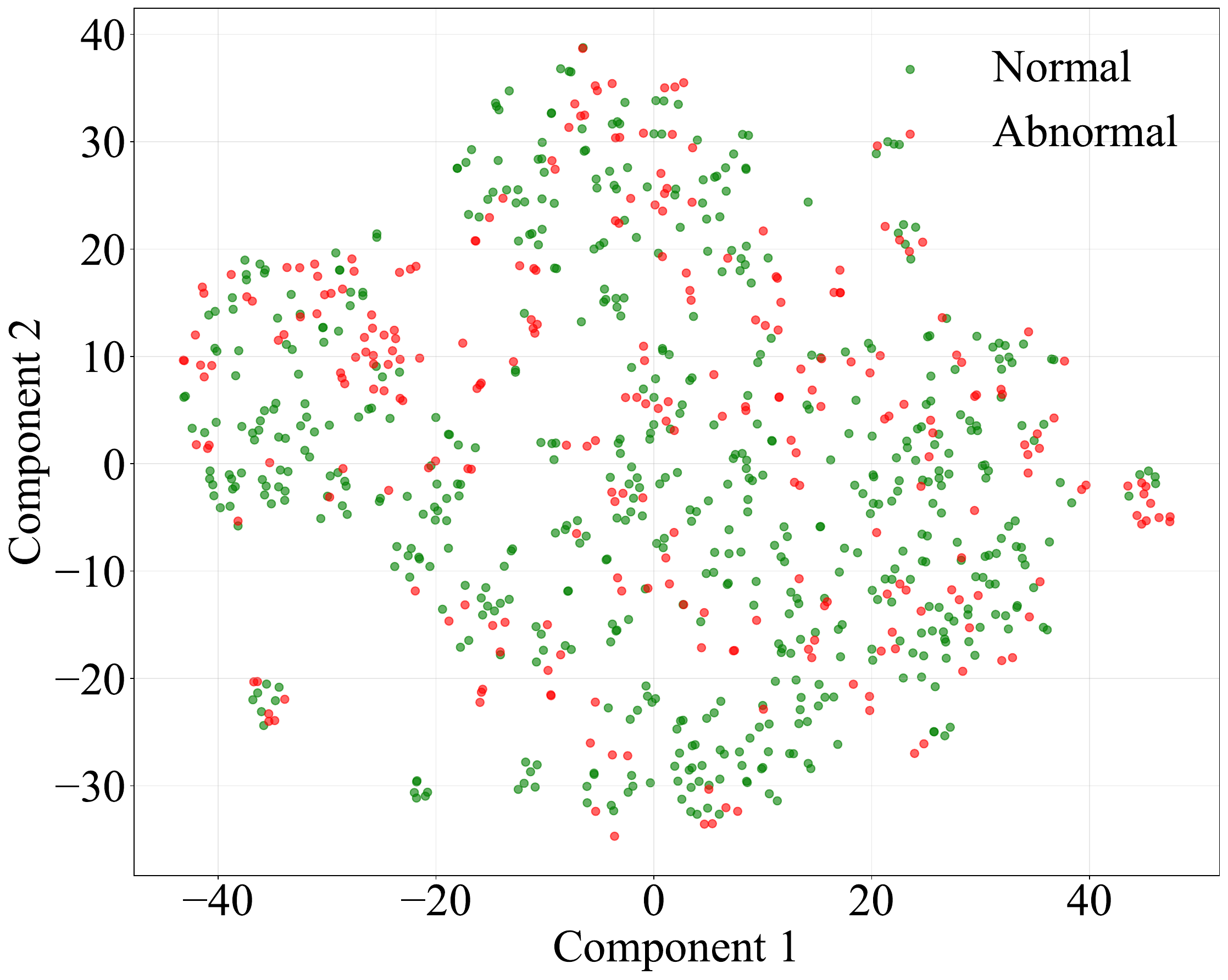}
			\label{fig: raw-model}
		\end{minipage}
	}
	\subfigure[RuntimeSlicer]{
		\begin{minipage}{0.47\linewidth}
			\centering
			\includegraphics[width=\textwidth]{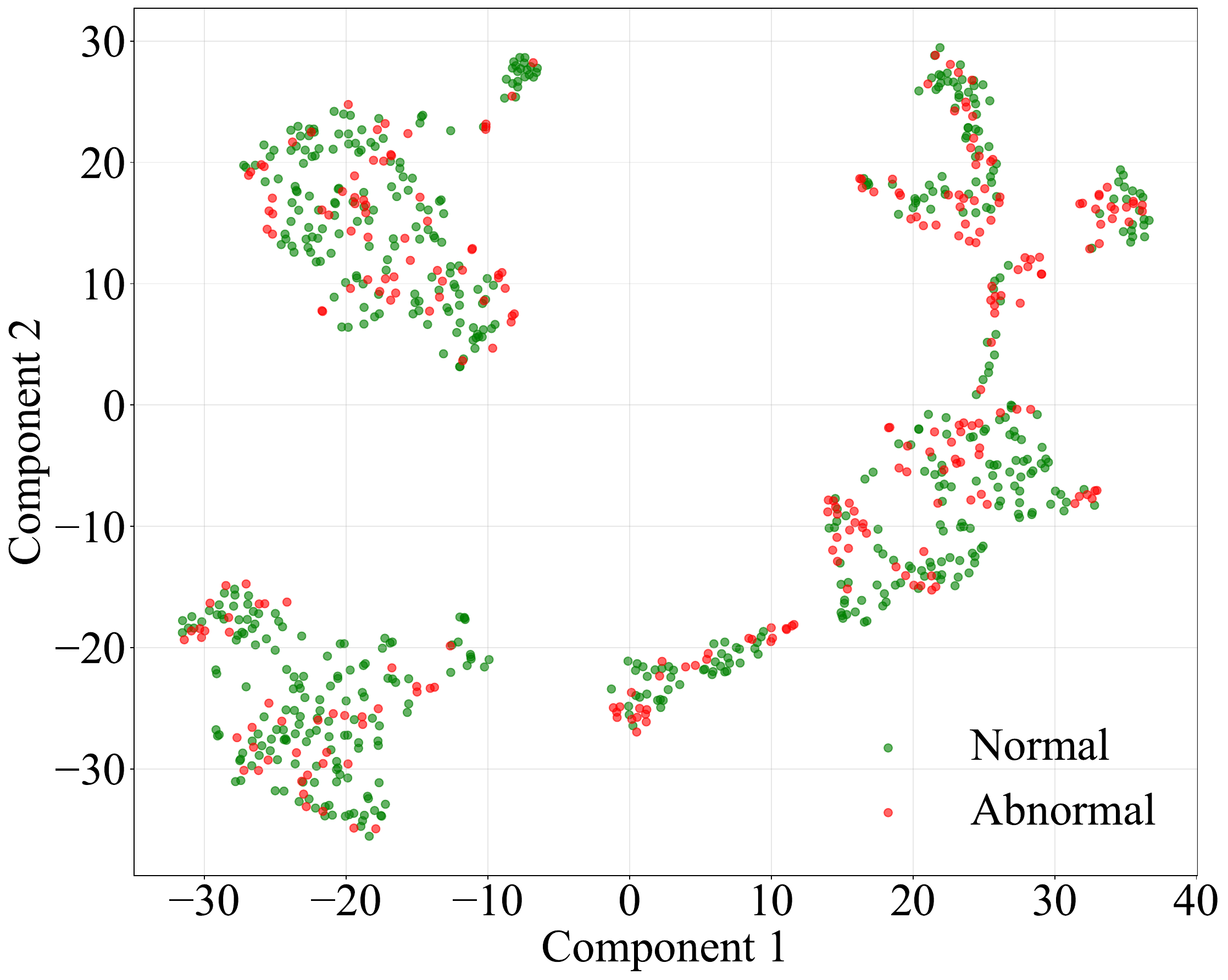}
			\label{fig: tunned-model}
		\end{minipage}
	}
	\vspace{-1.2em}
	\caption{Comparison of runtime-state embeddings generated by Qwen3-Embedding-0.6B and RuntimeSlicer}
	\label{fig: distinguish-evaluation}
	\vspace{-0.8em}
\end{figure}

We first examine the quality of RuntimeSlicer’s learned representations by evaluating its ability to distinguish runtime system states. As illustrated in Figure~\ref{fig: distinguish-evaluation}, we compare 2D t\text{-}SNE visualizations of runtime-state embeddings generated by Qwen3-Embedding-0.6B versus RuntimeSlicer. The baseline embeddings exhibit no clear structure, with points densely mixed across the space, whereas RuntimeSlicer yields well-separated clusters, indicating that it captures latent system-state patterns more effectively. This suggests that RuntimeSlicer provides state-aware representations that are more suitable for downstream operational tasks.

\vspace{-0.8em}

Next, we evaluate its performance on downstream failure management tasks. We report Precision, Recall, and F1 for all tasks; additionally, for Failure Localization and Failure Diagnosis, we also report Mean Reciprocal Rank (MRR). As shown in Table~\ref{tab: failure-management-result}, RuntimeSlicer combined with State-Aware Task-Oriented Tuning achieves promising performance across all three failure-management tasks. However, we observe performance degradation in cases where certain runtime states are underrepresented in the training data. We plan to focus on optimizing performance for such scenarios in future work.

\vspace{-0.8em}
\begin{table}[h]
	\centering
	\caption{Failure Management Results}
	\vspace{-0.8em}
	\label{tab: failure-management-result}
	\begin{tabular}{ccccc}
		\toprule
		\textbf{Task} & Precision & Recall & F1-Score & MRR \\
		\midrule
		Anomaly Detection & 97.27\% & 81.18\% & 88.50\%  & - \\
		Failure Localization & 69.57\% & 67.33\% & 68.43\%  & 70.15\% \\
		Failure Diagnosis & 87.57\% & 75.12\% & 80.88\% & 83.35\% \\
		\midrule
	\end{tabular}
	\vspace{-0.8em}
\end{table}

\section{Conclusion}

This paper presents a unified runtime state representation for generalizable failure management, eliminating the need for task-specific encoder training. To this end, we introduce Unified Runtime Contrastive Learning, which jointly leverages heterogeneous labeled and unlabeled data to learn system-state representations, yielding RuntimeSlicer. Based on these embeddings, we further propose a demonstration framework—State-Aware Task-Oriented Tuning—to showcase how RuntimeSlicer supports downstream failure-management tasks. Preliminary experiments verify its feasibility, and future work will focus on improving generalization through large-scale fine-tuning.

\section*{Acknowledgment}

This work was supported by the Huawei–Peking University Joint Laboratory of Mathematics.

\bibliographystyle{ACM-Reference-Format}
\balance
\bibliography{sample-base}

@String{Computing = "Computing" }

@String{Springer = "Springer-Verlag" }

@article{zhang2025survey,
	title={A Survey of AIOps in the Era of Large Language Models},
	author={Zhang, Lingzhe and Jia, Tong and Jia, Mengxi and Wu, Yifan and Liu, Aiwei and Yang, Yong and Wu, Zhonghai and Hu, Xuming and Yu, Philip and Li, Ying},
	journal={ACM Computing Surveys},
	year={2025},
	publisher={ACM New York, NY}
}

@inproceedings{zhang2025agentfm,
	title={Agentfm: Role-aware failure management for distributed databases with llm-driven multi-agents},
	author={Zhang, Lingzhe and Zhai, Yunpeng and Jia, Tong and Huang, Xiaosong and Duan, Chiming and Li, Ying},
	booktitle={Proceedings of the 33rd ACM International Conference on the Foundations of Software Engineering},
	pages={525--529},
	year={2025}
}

@inproceedings{zhang2025scalalog,
	title={ScalaLog: Scalable Log-Based Failure Diagnosis Using LLM},
	author={Zhang, Lingzhe and Jia, Tong and Jia, Mengxi and Wu, Yifan and Liu, Hongyi and Li, Ying},
	booktitle={ICASSP 2025-2025 IEEE International Conference on Acoustics, Speech and Signal Processing (ICASSP)},
	pages={1--5},
	year={2025},
	organization={IEEE}
}

@article{zhang2025thinkfl,
	title={ThinkFL: Self-Refining Failure Localization for Microservice Systems via Reinforcement Fine-Tuning},
	author={Zhang, Lingzhe and Zhai, Yunpeng and Jia, Tong and Duan, Chiming and Yu, Siyu and Gao, Jinyang and Ding, Bolin and Wu, Zhonghai and Li, Ying},
	journal={arXiv preprint arXiv:2504.18776},
	year={2025}
}

@inproceedings{zhang2025xraglog,
	title={XRAGLog: A Resource-Efficient and Context-Aware Log-Based Anomaly Detection Method Using Retrieval-Augmented Generation},
	author={Zhang, Lingzhe and Jia, Tong and Jia, Mengxi and Wu, Yifan and Liu, Hongyi and Li, Ying},
	booktitle={AAAI 2025 Workshop on Preventing and Detecting LLM Misinformation (PDLM)},
	year={2025}
}

@inproceedings{zhang2024multivariate,
	title={Multivariate log-based anomaly detection for distributed database},
	author={Zhang, Lingzhe and Jia, Tong and Jia, Mengxi and Li, Ying and Yang, Yong and Wu, Zhonghai},
	booktitle={Proceedings of the 30th ACM SIGKDD Conference on Knowledge Discovery and Data Mining},
	pages={4256--4267},
	year={2024}
}

@inproceedings{zhang2024reducing,
	title={Reducing events to augment log-based anomaly detection models: An empirical study},
	author={Zhang, Lingzhe and Jia, Tong and Wang, Kangjin and Jia, Mengxi and Yang, Yong and Li, Ying},
	booktitle={Proceedings of the 18th ACM/IEEE International Symposium on Empirical Software Engineering and Measurement},
	pages={538--548},
	year={2024}
}

@article{zhang2025log,
	title={E-log: Fine-grained elastic log-based anomaly detection and diagnosis for databases},
	author={Zhang, Lingzhe and Jia, Tong and Tan, Xinyu and Huang, Xiangdong and Jia, Mengxi and Liu, Hongyi and Wu, Zhonghai and Li, Ying},
	journal={IEEE Transactions on Services Computing},
	year={2025},
	publisher={IEEE}
}

@article{zhang2025adaptive,
	title={Adaptive root cause localization for microservice systems with multi-agent recursion-of-thought},
	author={Zhang, Lingzhe and Jia, Tong and Wang, Kangjin and Hong, Weijie and Duan, Chiming and He, Minghua and Li, Ying},
	journal={arXiv preprint arXiv:2508.20370},
	year={2025}
}

@inproceedings{liu2025ora,
	title={ORA: Job Runtime Prediction for High-Performance Computing Platforms Using the Online Retrieval-Augmented Language Model},
	author={Liu, Hongyi and Ma, Yinping and Huang, Xiaosong and Zhang, Lingzhe and Jia, Tong and Li, Ying},
	booktitle={Proceedings of the 39th ACM International Conference on Supercomputing},
	pages={884--894},
	year={2025}
}

@article{liu2024uac,
	title={Uac-ad: Unsupervised adversarial contrastive learning for anomaly detection on multi-modal data in microservice systems},
	author={Liu, Hongyi and Huang, Xiaosong and Jia, Mengxi and Jia, Tong and Han, Jing and Wu, Zhonghai and Li, Ying},
	journal={IEEE Transactions on Services Computing},
	volume={17},
	number={6},
	pages={3887--3900},
	year={2024},
	publisher={IEEE}
}

@article{he2025walk,
	title={Walk the talk: Is your log-based software reliability maintenance system really reliable?},
	author={He, Minghua and Jia, Tong and Duan, Chiming and Xiao, Pei and Zhang, Lingzhe and Wang, Kangjin and Wu, Yifan and Li, Ying and Huang, Gang},
	journal={arXiv preprint arXiv:2509.24352},
	year={2025}
}

@article{he2025united,
	title={United we stand: Towards end-to-end log-based fault diagnosis via interactive multi-task learning},
	author={He, Minghua and Duan, Chiming and Xiao, Pei and Jia, Tong and Yu, Siyu and Zhang, Lingzhe and Hong, Weijie and Han, Jin and Wu, Yifan and Li, Ying and others},
	journal={arXiv preprint arXiv:2509.24364},
	year={2025}
}

@inproceedings{liu2025aaad,
	title={AAAD: Asynchronous Inter-Variable Relationship-Aware Anomaly Detection for Multivariate Time Series},
	author={Liu, Hongyi and Huang, Xiaosong and Jia, Mengxi and Zhang, Lingzhe and Jia, Tong and Wu, Zhonghai and Li, Ying},
	booktitle={2025 IEEE International Conference on Multimedia and Expo (ICME)},
	pages={1--6},
	year={2025},
	organization={IEEE}
}

@inproceedings{hong2025cslparser,
	title={CSLParser: A Collaborative Framework Using Small and Large Language Models for Log Parsing},
	author={Hong, Weijie and Wu, Yifan and Zhang, Lingzhe and Duan, Chiming and Xiao, Pei and He, Minghua and Yang, Xixuan and Li, Ying},
	booktitle={2025 IEEE 36th International Symposium on Software Reliability Engineering (ISSRE)},
	pages={61--72},
	year={2025},
	organization={IEEE}
}

@article{huang2025uda,
	title={UDA-RCL: Unsupervised Domain Adaptation for Microservice Root Cause Localization Utilizing Multimodal Data},
	author={Huang, Xiaosong and Liu, Hongyi and Wu, Yifan and Zhang, Lingzhe and Jia, Tong and Li, Ying and Wu, Zhonghai},
	journal={IEEE Transactions on Services Computing},
	year={2025},
	publisher={IEEE}
}

@article{zhang2026hypothesize,
	title={Hypothesize-Then-Verify: Speculative Root Cause Analysis for Microservices with Pathwise Parallelism},
	author={Zhang, Lingzhe and Jia, Tong and Zhai, Yunpeng and Pan, Leyi and Duan, Chiming and He, Minghua and Xiao, Pei and Li, Ying},
	journal={arXiv preprint arXiv:2601.02736},
	year={2026}
}

@article{zhang2026agentic,
	title={Agentic Memory Enhanced Recursive Reasoning for Root Cause Localization in Microservices},
	author={Zhang, Lingzhe and Jia, Tong and Zhai, Yunpeng and Pan, Leyi and Duan, Chiming and He, Minghua and Jia, Mengxi and Li, Ying},
	journal={arXiv preprint arXiv:2601.02732},
	year={2026}
}

@article{zhang2025microremed,
	title={MicroRemed: Benchmarking LLMs in Microservices Remediation},
	author={Zhang, Lingzhe and Zhai, Yunpeng and Jia, Tong and Duan, Chiming and He, Minghua and Pan, Leyi and Liu, Zhaoyang and Ding, Bolin and Li, Ying},
	journal={arXiv preprint arXiv:2511.01166},
	year={2025}
}

@article{duan2025logaction,
	title={LogAction: Consistent Cross-system Anomaly Detection through Logs via Active Domain},
	author={Duan, Chiming and He, Minghua and Xiao, Pei and Jia, Tong and Zhang, Xin and Zhong, Zhewei and Luo, Xiang and Niu, Yan and Zhang, Lingzhe and Wu, Yifan and others},
	journal={arXiv preprint arXiv:2510.03288},
	year={2025}
}

@article{elliot2014devops,
	title={DevOps and the cost of downtime: Fortune 1000 best practice metrics quantified},
	author={Elliot, Stephen},
	journal={International Data Corporation (IDC)},
	year={2014}
}

@techreport{ITICCostofDowntime2024,
	author = {Information Technology Intelligence Consulting (ITIC)},
	title = {ITIC 2024 Global Server Hardware,Server OS Reliability Report},
	institution = {ITIC},
	year = {2024},
	type = {Annual Report}
}

@inproceedings{rasul2023lag,
	title={Lag-llama: Towards foundation models for time series forecasting},
	author={Rasul, Kashif and Ashok, Arjun and Williams, Andrew Robert and Khorasani, Arian and Adamopoulos, George and Bhagwatkar, Rishika and Bilo{\v{s}}, Marin and Ghonia, Hena and Hassen, Nadhir and Schneider, Anderson and others},
	booktitle={R0-FoMo: Robustness of Few-shot and Zero-shot Learning in Large Foundation Models},
	year={2023}
}

@article{dong2024simmtm,
	title={Simmtm: A simple pre-training framework for masked time-series modeling},
	author={Dong, Jiaxiang and Wu, Haixu and Zhang, Haoran and Zhang, Li and Wang, Jianmin and Long, Mingsheng},
	journal={Advances in Neural Information Processing Systems},
	volume={36},
	year={2024}
}

@article{liao2025timegpt,
	title={TimeGPT in load forecasting: A large time series model perspective},
	author={Liao, Wenlong and Wang, Shouxiang and Yang, Dechang and Yang, Zhe and Fang, Jiannong and Rehtanz, Christian and Port{\'e}-Agel, Fernando},
	journal={Applied Energy},
	volume={379},
	pages={124973},
	year={2025},
	publisher={Elsevier}
}

@inproceedings{zhang2024trace,
	title={Trace-based Multi-Dimensional Root Cause Localization of Performance Issues in Microservice Systems},
	author={Zhang, Chenxi and Dong, Zhen and Peng, Xin and Zhang, Bicheng and Chen, Miao},
	booktitle={Proceedings of the IEEE/ACM 46th International Conference on Software Engineering},
	pages={1--12},
	year={2024}
}

@inproceedings{yu2021microrank,
	title={Microrank: End-to-end latency issue localization with extended spectrum analysis in microservice environments},
	author={Yu, Guangba and Chen, Pengfei and Chen, Hongyang and Guan, Zijie and Huang, Zicheng and Jing, Linxiao and Weng, Tianjun and Sun, Xinmeng and Li, Xiaoyun},
	booktitle={Proceedings of the Web Conference 2021},
	pages={3087--3098},
	year={2021}
}

@article{li2022swisslog,
	title={SwissLog: Robust anomaly detection and localization for interleaved unstructured logs},
	author={Li, Xiaoyun and Chen, Pengfei and Jing, Linxiao and He, Zilong and Yu, Guangba},
	journal={IEEE Transactions on Dependable and Secure Computing},
	volume={20},
	number={4},
	pages={2762--2780},
	year={2022},
	publisher={IEEE}
}

@article{sui2023logkg,
	title={Logkg: Log failure diagnosis through knowledge graph},
	author={Sui, Yicheng and Zhang, Yuzhe and Sun, Jianjun and Xu, Ting and Zhang, Shenglin and Li, Zhengdan and Sun, Yongqian and Guo, Fangrui and Shen, Junyu and Zhang, Yuzhi and others},
	journal={IEEE Transactions on Services Computing},
	volume={16},
	number={5},
	pages={3493--3507},
	year={2023},
	publisher={IEEE}
}

@inproceedings{li2022microsketch,
	title={MicroSketch: Lightweight and adaptive sketch based performance issue detection and localization in microservice systems},
	author={Li, Yufeng and Yu, Guangba and Chen, Pengfei and Zhang, Chuanfu and Zheng, Zibin},
	booktitle={International Conference on Service-Oriented Computing},
	pages={219--236},
	year={2022},
	organization={Springer}
}

@inproceedings{gan2021sage,
	title={Sage: practical and scalable ML-driven performance debugging in microservices},
	author={Gan, Yu and Liang, Mingyu and Dev, Sundar and Lo, David and Delimitrou, Christina},
	booktitle={Proceedings of the 26th ACM International Conference on Architectural Support for Programming Languages and Operating Systems},
	pages={135--151},
	year={2021}
}

@inproceedings{cai2021tracemodel,
	title={Tracemodel: An automatic anomaly detection and root cause localization framework for microservice systems},
	author={Cai, Yang and Han, Biao and Su, Jinshu and Wang, Xiaoyan},
	booktitle={2021 17th International Conference on Mobility, Sensing and Networking (MSN)},
	pages={512--519},
	year={2021},
	organization={IEEE}
}

@inproceedings{zhang2022crisp,
	title={CRISP: Critical path analysis of Large-Scale microservice architectures},
	author={Zhang, Zhizhou and Ramanathan, Murali Krishna and Raj, Prithvi and Parwal, Abhishek and Sherwood, Timothy and Chabbi, Milind},
	booktitle={2022 USENIX Annual Technical Conference (USENIX ATC 22)},
	pages={655--672},
	year={2022}
}

@inproceedings{lee2023eadro,
	title={Eadro: An end-to-end troubleshooting framework for microservices on multi-source data},
	author={Lee, Cheryl and Yang, Tianyi and Chen, Zhuangbin and Su, Yuxin and Lyu, Michael R},
	booktitle={2023 IEEE/ACM 45th International Conference on Software Engineering (ICSE)},
	pages={1750--1762},
	year={2023},
	organization={IEEE}
}

@article{zhang2023robust,
	title={Robust failure diagnosis of microservice system through multimodal data},
	author={Zhang, Shenglin and Jin, Pengxiang and Lin, Zihan and Sun, Yongqian and Zhang, Bicheng and Xia, Sibo and Li, Zhengdan and Zhong, Zhenyu and Ma, Minghua and Jin, Wa and others},
	journal={IEEE Transactions on Services Computing},
	volume={16},
	number={6},
	pages={3851--3864},
	year={2023},
	publisher={IEEE}
}

@inproceedings{yu2023nezha,
	title={Nezha: Interpretable fine-grained root causes analysis for microservices on multi-modal observability data},
	author={Yu, Guangba and Chen, Pengfei and Li, Yufeng and Chen, Hongyang and Li, Xiaoyun and Zheng, Zibin},
	booktitle={Proceedings of the 31st ACM Joint European Software Engineering Conference and Symposium on the Foundations of Software Engineering},
	pages={553--565},
	year={2023}
}

@article{sun2025interpretable,
	title={Interpretable failure localization for microservice systems based on graph autoencoder},
	author={Sun, Yongqian and Lin, Zihan and Shi, Binpeng and Zhang, Shenglin and Ma, Shiyu and Jin, Pengxiang and Zhong, Zhenyu and Pan, Lemeng and Guo, Yicheng and Pei, Dan},
	journal={ACM Transactions on Software Engineering and Methodology},
	volume={34},
	number={2},
	pages={1--28},
	year={2025},
	publisher={ACM New York, NY}
}

@inproceedings{lin2024root,
	title={Root Cause Analysis in Microservice Using Neural Granger Causal Discovery},
	author={Lin, Cheng-Ming and Chang, Ching and Wang, Wei-Yao and Wang, Kuang-Da and Peng, Wen-Chih},
	booktitle={Proceedings of the AAAI Conference on Artificial Intelligence},
	volume={38},
	number={1},
	pages={206--213},
	year={2024}
}

@article{yu2023tracerank,
	title={TraceRank: Abnormal service localization with dis-aggregated end-to-end tracing data in cloud native systems},
	author={Yu, Guangba and Huang, Zicheng and Chen, Pengfei},
	journal={Journal of Software: Evolution and Process},
	volume={35},
	number={10},
	pages={e2413},
	year={2023},
	publisher={Wiley Online Library}
}

@misc{aiops2022championship,
	title        = {{AIOPS 2022 Championship}},
	howpublished = {\url{https://competition.aiops.cn/}},
	year         = {2022}
}

@inproceedings{sun2024art,
	title={Art: A unified unsupervised framework for incident management in microservice systems},
	author={Sun, Yongqian and Shi, Binpeng and Mao, Mingyu and Ma, Minghua and Xia, Sibo and Zhang, Shenglin and Pei, Dan},
	booktitle={Proceedings of the 39th IEEE/ACM International Conference on Automated Software Engineering},
	pages={1183--1194},
	year={2024}
}

@article{zhou2018fault,
	title={Fault analysis and debugging of microservice systems: Industrial survey, benchmark system, and empirical study},
	author={Zhou, Xiang and Peng, Xin and Xie, Tao and Sun, Jun and Ji, Chao and Li, Wenhai and Ding, Dan},
	journal={IEEE Transactions on Software Engineering},
	volume={47},
	number={2},
	pages={243--260},
	year={2018},
	publisher={IEEE}
}

@misc{google2025onlineboutique,
	title        = {Online Boutique: A Cloud-First Microservices Demo Application},
	author       = {{Google Cloud Platform}},
	year         = {2025},
	howpublished = {\url{https://github.com/GoogleCloudPlatform/microservices-demo}},
	note         = {Accessed: October 15, 2025}
}

\clearpage

\end{sloppypar}
\end{document}